\documentclass[a4paper,12pt]{article}

\usepackage[utf8]{inputenc}
\usepackage{amsmath}
\usepackage{pdflscape}
\usepackage{amsfonts}
\usepackage{amssymb}
\usepackage{natbib}
\usepackage{graphicx}
\usepackage{textcomp}
\usepackage{color}
\usepackage[breaklinks]{hyperref}
\usepackage{titlesec} 
\usepackage[nottoc,notlot,notlof]{tocbibind}
\usepackage[letterpaper,top=3.2cm, left=2.5cm, right=3cm, bottom=0.8cm]{geometry}
\batchmode

\newcommand{\Rea}{\mathbb{R}}

\newcommand{\pei}{\langle}
\newcommand{\ped}{\rangle}

\usepackage{authblk}

\title{Robust clustering for functional data based on trimming and constraints}
\author[a]{Rivera-Garc\'{\i}a D.}
\author[b]{Garc\'{\i}a-Escudero L.A.}
\author[b]{Mayo-Iscar A.}
\author[a]{Ortega, J.}
\affil[a]{CIMAT, A.C. Jalisco s/n, Mineral de Valenciana. Guanajuato 36240, Mexico.}
\affil[b]{Dept. de Estad\'istica e Investigaci\'on Operativa, Universidad de Valladolid. Paseo de Bel\'en, 7. 47005 Valladolid. Spain.}

\date{}

\begin{document}
\bibliographystyle{chicago}

\maketitle

%

\section*{Abstract}
Many clustering algorithms when the data are curves or functions
have been recently proposed. However, the presence of contamination
in the sample of curves can influence the performance of most of them.
In this work we propose a robust, model-based
clustering method based on an approximation to the ``density
function" for functional data. The robustness results from the joint
application of trimming, for reducing the effect of contaminated
observations, and constraints on the variances, for avoiding
spurious clusters in the solution. The proposed method has been
evaluated through a simulation study. Finally, an application to a
real data problem is given.

{\bf Keyworks:} Functional data analysis $\cdot$ clustering $\cdot$
robustness $\cdot$ functional principal components analysis.

\section{Introduction}
Recent technological advances have provided more precise
instruments, which make possible the recording of large numbers of
subsequent measurements in such a way that data can be considered as
realizations of random continuous functions. In this context,
Functional Data Analysis \citep{RS2006,FV2006} has received
increasing attention in recent years. Cluster analysis consists of
identifying homogeneous groups within a data set and there is also a
need for appropriate clustering methods for this new type of
functional data sets.

There are many methods to perform cluster analysis for traditional
multivariate data among which stand several based on
probabilistic models (model-based clustering). The use of the EM
algorithm is quite common in order to solve the likelihood
maximization involved in all these approaches \citep{Fraley2002}.
Posterior probabilities are used  to estimate the probabilities of
membership of an observation to a specific group.

Several clustering methods for functional data have been recently
proposed. A first approximation is known as raw-data clustering,
which consists of using the discretization of the functions
and directly applying traditional multivariate clustering
techniques. A second approximation is based on a reconstruction of
the functional form of the data through the use of basis of
functions such as B-splines, Wavelets, Fourier series, etc. \citep{RS2006}.
in this case, usual clustering techniques are applied to the fitted
coefficients for the functional representation of  each curve.
Another approach is based on probabilistic models, where
a probability distribution for those coefficients is assumed as, for instance in
\cite{jamessugar2003} and more recently \cite{JP2013}, who employ an
approximation to the ``density function" for functional data proposed in
\citep{DH2010}.

However, the determination of an appropriate clustering technique is
even more difficult under the possible presence of outlying curves.
One possibility to robustify clustering algorithms is through the
application of trimming tools (\cite{CGmayo1997},
\cite{gallegos2002}, \cite{gallegos2002}). In \cite{GEV2008},
\cite{GEV2014}, \cite{Escudero2015} and \cite{Fritz2013},
restrictions on the matrices of dispersion of the groups are also
introduced to avoid the detection of spurious clusters.

Trimming techniques have been already applied as a robust functional
clustering tool (\cite{Efuncional2005} and
\cite{cuestafraiman2007}). This work provides an extension of these
principles but in a more model-based approach.

The outline for the rest of this work is as follows. In Section \ref{se_2}, we
give a brief description of the approximation to the ``density" for
functional data that will be applied later. A model-based clustering
for functional data is presented in Section \ref{se_3}. Our proposal
for robust functional clustering (RFC) and a feasible algorithm for it are
described in Section \ref{se_4}. Sections \ref{Simustudy} and
\ref{real_data} present a simulation study and real data example to
illustrate the performance of the proposed methodology. Finally, we
give our conclusions in Section \ref{conclusions}.

\section{Approximation to the ``density function" for functional data}\label{se_2}
Let $L^{2}([0,T])$ be a Hilbert space of functions with inner
product given by $\pei f,g \ped= \int f(t)g(t)\, dt$  and norm $||\cdot
||=\pei \cdot ,\cdot \ped^{1/2}$. Suppose $X$ is a random function
in $L^{2}([0,T])$. Assume the process $X$ has mean
$\mu(t)=E\{X(t)\}$ and covariance
$\Gamma(s,t)=\text{cov}\{X(s),X(t)\}$ which are smooth
continuous functions. Consider the Karhunen-Lo\`eve (K-L) expansion:
$$
 X(t)=\mu(t)+\sum_{j=1}^{\infty}C_{j}(X)\psi_{j}(t)
$$
where the eigenfunctions $\psi_{j}$ form an orthonormal system and
are associated with the covariance operator $\Gamma$ by means of the
corresponding eigenvalues $\lambda_{j}$ so that $\pei
\Gamma(\cdot,t) ,\psi_{j} \ped=\lambda_{j}\psi_{j}(t)$. The
eigenfunctions are orthogonal, i.e. they satisfy $\pei \psi_{l}, \psi_{k}
\ped=\delta_{lk}$, where $\delta_{lk}$ is the Kronecker delta with
$1$ if $l=k$ and $0$ otherwise. The eigenvalues are
assumed to be in decreasing order, $\lambda_{1}\geq\lambda_{2}\geq
\cdots,$ with $\sum_{j=1}^{\infty}\lambda_{j}<\infty$. The
coefficients $C_{j}(X)$, $j=1,2,...$, better known as principal
components or scores of the stochastic process $X$, are uncorrelated
random variables with zero mean and variance $\lambda_{j}$, such
that $ C_{j}(X)=\pei X-\mu, \psi_{j}\ped$ is the projection of
$X-\mu$ on the $j$-th eigenfunction $\psi_{j}$.

Let $X^{(p)}$ be the approximation of $X$ based
on the $p$ first terms in the K-L expansion, this is
\begin{align}\label{eq3}
 X^{(p)}(t)=\mu(t)+\sum_{j=1}^{p}C_{j}(X)\psi_{j}(t).
\end{align}
It can be seen that $E(||X-X^{(p)}||^{2})=\sum_{j\geq
p+1}\lambda_{j}\;\;\;\;\; \text{and} \;\;\;\;\;
||X-X^{(p)}||\xrightarrow {m.s.}0$ when $p \rightarrow \infty$.

The notion of density for functional data is not well defined, but
there exist some approximations to the density function in the
literature. For example, \cite{FV2006} have developed extensions of
the multivariate case in the nonparametric context.

Without loss of generality, let us suppose that $X$ is a zero mean
stochastic process, i.e. $\mu(t)=0$ for every $t \in [0,T]$ which
can be approximated by $ X^{(p)}$ as in (\ref{eq3}). According to
\cite{DH2010}, it is possible to approximate the ``density function"
for functional data $X$, when functions are considered in the space
determined by the eigenfunctions of the principal components. The
notion of small ball probability has an important role in the
development of the approximation. Based on the K-L expansion
$X^{(p)}$, \cite{DH2010} show that the probability that $X$ belongs
to a ball of radius $h$ centred in $x\in L_{2}[0,T]$ can be written
as
$$
\log P(||X-x||\leq h)=\sum_{j=1}^{p}\log
f_{C_{j}}(c_{j}(x))+\xi(h,\rho(h))+o(\rho(h))
$$
where $||X-x||$ denotes the $L_2$ distance between $X$ and $x$,
$f_{C_{j}}$ corresponds to the probability density of $C_{j}$ and
$c_{j}(x)=\pei x, \psi_{j} \ped_{L_{2}}$ is the $j$-th principal
component or score of $x$. $\rho$ and $\xi$ are functions such that
$\rho$ increases to infinity when $h$ decreases to zero. Note that
$\log P(||X-x||\leq h)$ depends on $x$ through the term
$\sum_{j=1}^{p}\log f_{C_{j}}(c_{j}(x))$. This term captures the
first-order effect that $x$ has on $\log P(||X-x||\leq h)$.
Therefore, it serves to describe the main differences in  sizes of
small-ball probabilities for different values of $x$ since the
notion of probability density in the finite dimensional case can be
seen as the limit of $P(||X-x||\leq h)/h$ when $h$ tends to zero.
Moreover, as seen in \cite{JP2013}, it can be observed that for
every $h>0$ and $x\in L_{2}([0,T])$,
\begin{align}
 P(||X^{(p)}-x||\leq h-||X-X^{(p)}||)\leq P(||X-x||\leq h)\leq P(||X^{(p)}-x||\leq
 h+||X-X^{(p)}||).
\end{align}
Hence, the probability $P(||X-x||\leq h)$ can be approximated by
$P(||X^{(p)}-x||\leq h)$. If $f_{X}^{(p)}$ is the joint density
function of $C^{(p)}=(C_{1},...,C_{p})$ and
$x=\sum_{j\geq1}c_{j}(x)\psi_{j}$
 then
$$
 P(||X^{(p)}-x||\leq h)=\int_{\mathcal{D}_{x}^{(p)}}f_{X}^{(p)}(y)\, dy,
$$
with $x^{(p)}=\sum_{j=1}^{p}c_{j}(x)\psi_{j}$ and
${\mathcal{D}_{x}^{(p)}}=\left\{y\in
\Rea^{p}:||y-x^{(p)}||_{\Rea^{p}}\leq\sqrt{h^{2}-\sum_{j\geq
p+1}c_{j}^{2}(x)}\right\}$. In this way, the density of
$f_{X}^{(p)}$ can be seen as an approximation of the density of $X$.

Finally, we can also take into account that the principal components
$C_{j}$ are independent Gaussian random variables when $X$
corresponds to a Gaussian process. In this particular case,
$f^{(p)}_{X}$ is given by
$$
 f_{X}^{(p)}(x)=\prod_{j=1}^{p}f_{C_{j}}(c_{j}(x))
$$
where $f_{C_{j}}$ is a Gaussian density function with zero mean
and variance $\lambda_j$.

\section{Model-based clustering for functional data}\label{se_3}

In a clustering framework, we will consider $K$ different models,
one for each group. Conditional on the group $g$, let us consider
K-L expansions with $p$ terms and the density function approximation
as reviewed in Section \ref{se_2}, such that the density of the
principal components in groups are assumed uncorrelated Gaussian
variables with zero mean. Moreover, in order to simplify the largely
parameterized problem appearing in these clustering frameworks, we
consider that the first $q_g$ terms have no restrictions, as in
\cite{JP2013}, while the remaining $p-q_g$ are constrained in such a
way that their corresponding principal components have equal
variances, as done in \cite{BJ2011}. In other words, we assume that
scores in each group can be approximated by means of uncorrelated
random Gaussian variables with zero mean and covariance matrix
$\Sigma_{g}=\text{diag}(a_{1g},...,a_{q_gg},b_{g},...,b_{g})$ with
$a_{jg}>b_{g}$. This means that the main variances of the $g$-th
group are modeled by $a_{1g}$,..., $a_{q_gg}$ while $b_{g}$ serves
to model the variance of the noise of the residual process. If
$Z_{g}$ is a random indicator variable designating membership to
group $g$, for $g=1,2,..,K$, then we assume
$$
 f_{X\mid Z_{g}=1}^{q_g}(x)=\prod_{j=1}^{q_{g}}f_{C_{j}\mid Z_{g}=1}(c_{jg}(x);a_{jg})\prod_{j=q_{g}+1}^{p}f_{C_{j}\mid
 Z_{g}=1}(c_{jg}(x);b_{g}).
$$
Note that when $q_g=p$ we have the model proposed by \cite{JP2013}.

Assume now that $Z=(Z_{1},...,Z_{K})$ have a multinomial
distribution $\mathcal{M}(\pi_{1},...,\pi_{K})$, where
$\pi_{1},...,\pi_{K}$ are the mixture probabilities weights (with
$\sum_{g=1}^{K}\pi_{g}=1$). In this way, the unconditional
``approximated functional density" of $X$ at $x\in L^2([0,T])$ is
given by
$$
 f_{X}(x;\theta)=\sum_{g=1}^{K} \pi_{g}\left[ \prod_{j=1}^{q_{g}}f_{C_{j}\mid Z_{g}=1}(c_{jg}(x);a_{j,g})\prod_{j=q_{g}+1}^{p}f_{C_{j}\mid
 Z_{g}=1}(c_{jg}(x);b_{g})\right],
$$
where $\theta$ denotes all the parameters that need to be estimated
in that expression. Notice that, to start, we are assuming that
$(q_1,...,q_K)$ are known in advance dimensions.

Suppose now that $\{x_1,...,x_n\}$ is a set of curves being the
realization from an independent, identically distributed (i.i.d.)
sample from $X\in \mathcal{L}_2([0,T])$. We define the
mixture-loglikelihood function as
$$
 l^{p}(\theta;x_1,...,x_n)=\sum_{i=1}^{n} \log \left(\sum_{g=1}^{K} \pi_{g} \left[\prod_{j=1}^{q_{g}} \frac{1}{\sqrt{2 \pi a_{jg}}} \exp\left(\frac{-c_{ijg}^{2}}{2a_{jg}}\right)\prod_{j=q_{g}+1}^{p} \frac{1}{\sqrt{2 \pi b_{g}}} \exp\left(\frac{-c_{ijg}^{2}}{2 b_{g}}\right)\right]\right),
$$
where $c_{ijg}=c_{jg}(x_{i})$ corresponds to the $j$-th principal
component of the curve $x_{i}$ in group $g$.

\section{Robust functional clustering based on trimming and constraints}\label{se_4}

\cite{GEV2014} and \cite{RG2015} provide a methodology for robust
mixture modeling in a multivariate real-valued context. This
methodology proposes using trimming and scatter constraints to
remove the contamination in the data and simultaneously avoid
spurious clusters. The methodology is also based on a particular
type of trimming which is determined by the dataset itself. To be
more precise, if $\{x_{1},...,x_{n}\}$ is a random sample in
$\Rea^{p}$, the idea is to maximize the trimmed mixture
likelihood defined as
\begin{align}\label{spurio}
 \sum_{i=1}^{n}
 \eta(x_{i})\log\left[\sum_{g=1}^{K}\pi_{g}\phi(x_{i};\theta_{g})\right],
\end{align}
where $\phi(\cdot;\theta_{g})$ stands for the $p$-dimensional
Gaussian density with parameters $\theta_{g}=(\mu_{g},\Sigma_{g})$.
The indicator function $\eta(\cdot)$ serves to designate whether the
observation $x_{i}$ has been trimmed (when $\eta(x_{i})=0$) or not
(when $\eta(x_{i})=1$). A proportion $\alpha$ of observations is
trimmed, so that $\sum_{i=1}^{n}\eta(x_{i})=[n(1-\alpha)]$.
Constraints on the eigenvalues of the scatter matrices are also
applied in order to make the maximization problem well defined and
to avoid the detection of non-interesting spurious solutions
(\cite{GEV2008}). Then, it is proposed to maximize \eqref{spurio}
subject to the restriction
$$
 \frac{\max_{g,j}\lambda_{j}(\Sigma_{g})}{\min_{g,j}\lambda_{j}(\Sigma_{g})}\leq
 d,
$$
where $\{\lambda_{j}(\Sigma_{g})\}_{j=1}^p$ is the eigenvalue set
for matrix $\Sigma_{g}$ and $d\geq1$ is a fixed constant. In the
most constrained case ($d=1$), we are searching for homoscedastic and
spherical clusters.

In a similar fashion, we can adapt this methodology for functional
data by considering a trimmed and constrained version of the
model-based clustering approach presented in Section \ref{se_3}. Let
$\{x_1,...,x_n\}$ be a realization from a i.i.d. sample of the
process $X \in L^{2}[0,T]$. A trimmed loglikelihood can be defined
in this functional setting as
\begin{align}\label{eq11}
 l^{p}_{\alpha}(\theta;x_1,...,x_n)=\sum_{i=1}^{n} \eta(x_{i}) \log \left(\sum_{g=1}^{K} \pi_{g} \left[\prod_{j=1}^{q_{g}} \frac{1}{\sqrt{2 \pi a_{jg}}} \exp\left(\frac{-c_{ijg}^{2}}{2a_{jg}}\right)\prod_{j=q_{g}+1}^{p} \frac{1}{\sqrt{2 \pi b_{g}}} \exp\left(\frac{-c_{ijg}^{2}}{2 b_{g}}\right)\right]\right)
\end{align}
where $c_{ijg}=c_{jg}(x_{i})$ is the $j$-th principal component
corresponding to curve $x_{i}$ in group $g$ and, again,
$\sum_{i=1}^{n}\eta(x_{i})=[n(1-\alpha)]$. To avoid spurious
solutions, we set two constants $d_1$ and $d_2$, both greater or
equal than 1, and impose the following constraints on the scatter
parameters:
$$
\frac{\max_{g=1,...,K;j=1,...,q_j}a_{jg}}{\min_{g=1,...,K;j=1,...,q_j}a_{jg}}\leq
d_1$$
and
$$\frac{\max_{g=1,...,K}b_{g}}{\min_{g=1,...,K}b_{g}}\leq d_2.
$$

\subsection{Proposed algorithm}\label{se_5}

Of course, the maximization of the trimmed log-likelihood in
(\ref{eq11}) may not be an easy task from a computational point
of view. A classical way of maximizing mixture model likelihoods is
to use the EM algorithm. The algorithm proposed here is based on the
traditional EM algorithm incorporating some additional steps. In
a so-called T-step (Trimming step) we temporally discard those
observations with smallest contributions to the likelihood (to
increase as much as possible the trimmed log-likelihood). We also
consider, in the M-step, a final refinement where the required
constraints on the scatter matrices are imposed on the scatter
parameters.

The proposed algorithm may be described as follows, where
$\theta^{(l)}$ are the values of parameters at stage $l$ of the
iterative process:

\begin{enumerate}
  \item[1.] \textit{Initialization:}
  The algorithm is randomly initialized \texttt{nstart} times by selecting
  different starting $\theta^{(0)}$ parameters.  With this idea in mind,
  we simply propose to randomly select $K\times h$ subindexes $\{i_{g_1},i_{g_2},...,i_{g_h}\}_{g=1}^K\subset \{1,2,...,n\}$ where $h$ is the minimum
  number of observations needed to computationally carry out a
  functional principal component analysis for these observations. We then
  apply the procedure
  that will be latter described in Step 2 of this algorithm with weights $\tau_{i_{g_1}g}=\tau_{i_{g_2}g}=....=\tau_{i_{g_h}g}=1$,
  $g=1,...,K$, and weights $\tau_{ig}=0$ for all the remaining $(i,g)$ pairs. The smaller the $h$ the more
  likely is that these  $K\times h$ observations could be
  free of outliers (or at least with not so many within) in any of those random initializations.

  \item[2.] \textit{Trimmed EM steps:}
  The following steps are alternatively executed until convergence
  (i.e. $\theta^{(l+1)}=\theta^{(l)}$) or a maximum number of iterations
  \texttt{iter.max} is reached.

  \begin{itemize}
    \item[2.1.] \textit{T- and E-steps:} Let us use the notation
    $$D_g(x_i,\theta)=\pi_{g} \prod_{j=1}^{q_{g}} \frac{1}{\sqrt{2 \pi a_{jg}}} \exp\left(\frac{-c_{ijg}^{2}}{2a_{jg}}\right)\prod_{j=q_{g}+1}^{p} \frac{1}{\sqrt{2 \pi b_{g}}} \exp\left(\frac{-c_{ijg}^{2}}{2 b_{g}}\right)$$
    and $$D(x_i,\theta)=\sum_{g=1}^K D_g(x_i,\theta).$$
    If we consider
    $
    D(x_{(1)};\theta^{(l)})\leq D(x_{(2)};\theta^{(l)}) \leq.... \leq
    D(x_{(n)};\theta^{(l)}),
    $
    the observations with indexes in
    \begin{equation}\label{i}
         I=\{i : D(x_i;\theta^{(l)})\leq D(x_{([n\alpha])};\theta^{(l)}) \}
    \end{equation}
    are those which are tentatively discarded in this iteration of the
    algorithm.

    As in other mixture
    fitting EM algorithms, we compute posterior
    probabilities by using the well-known Bayes rule as
    $$
        \tau_g(x_i;\theta^{(l)})=
        D_g(x_i;\theta^{(l)})/D(x_i;\theta^{(l)}), \text{ for }i=1,...,n.
    $$
    However, unlike standard EM algorithms, the
    $\tau_g(x_i;\theta^{(l)})$ values for the discarded
    observations are modified as
    $$
    \tau_g(x_i;\theta^{(l)})=0,\text{ for all }g=1,...,K,\text{ when }i\in
    I.
    $$
    Notice that the way that trimming is done is
  similar to that in \cite{GEV2014}.

    \item[2.2.] \textit{M-step:} This step consists of three stages:

    \begin{itemize}
      \item[2.2.1] \textit{Weights update:} Weights are updated as
      $$
    \pi_g^{(l+1)}= \sum_{i=1}^n \tau_g(x_i;\theta^{(l)})/[n(1-\alpha)]
    $$
      \item[2.2.2] \textit{Principal component update:} Consider a basis of functions
      $\Phi=\{\phi_{1},...,\phi_{p}\}$. If
      $x_i$ admits an approximate reconstruction in this basis as $x_{i}(t)\simeq \sum_{j=1}^{p}\gamma_{ij}\phi_{j}(t)$
      then let $\Gamma$ be the $n \times p$ matrix of coefficients $ \gamma_ {ij} $
      used in that reconstruction. Let $W$ be the matrix of the inner products between the basis functions
      $ W_{jl} = \int_{0} ^ {T} \phi_{j}(t) \phi_{l} (t) dt\; (1 \leq j, l \leq p) $.
      The updating of the principal components  is carried out by weighting
      the importance of the untrimmed $x_{i}(t)$ curves by the conditional probability $T_ {g}^{(l)}= \text{diag} (\tau_ {1, g}^{(l)}, ..., \tau_ {n, g}^{(l)})$.
      The first step is  to center the curve $ x_ {i} (t) $ in group $g$, by subtracting
      the weighted pointwise sample mean calculated with $ \tau_ {i, g}^{(l)}$ weights.
      The expansion coefficients of the centered curves are given by
      $ \Gamma_ {g}^{(l)} = (I_ {n}  -1_{n}(\tau_ {1, g}^{(l)}, ..., \tau_ {n, g}^{(l)})) \Gamma,$
      where $I_n$ is the $n\times n$ identity matrix and $1_n=(1,1,...,1)$ is the unit vector. Note
      that the weighted sample covariance function is then given by
      \begin{equation}\label{ec5}
       v^{(l+1)} (s, t) = \frac {1} {n _ {g}^{(l)}} \sum_ {i = 1} ^ {n} \tau_ {ig}^{(l)}
      x_ {i} (s) x_ {i} (t),
      \end{equation}
      where $n _ {g}^{(l)} = \sum_ {i = 1} ^ {n} \tau_ {ig}^{(l)}.$
      Consider also that the $j$-th eigenfunction can be written as $\psi_{j}(s) = \beta_{j} ^ {T} \phi
      (s)$ with $\phi(t)=(\phi_1(t),...,\phi_p(t))'$.
      Substituting
      the above expressions into (\ref{ec5}) 
      one obtains
      $$
      v^{(l+1)} (s, t) = (n _ {g} ^ {(l)})^{-1} \phi ^ {T} (s) \Gamma_ {g} ^ {T(l)} T_{g}^{(l)} \Gamma_ {g}^{(l)} \phi (t),
      $$
      and one gets the eigenequation
      $$
      (n _ {g} ^ {(l)})^{-1} \phi ^ {T} (s) \Gamma_ {g} ^ {T} T_ {g}^{(l)} \Gamma_{g} \left(\int \phi (t) \phi ^ {T} (t) dt \right) \beta_ {j} = \lambda \phi ^ {T} (s)\beta_ {j}.
      $$
      By using the $W$ matrix of the inner products, the previous
      equation can be written as
      $$(n _ {g} ^ {(l)})^{-1} \phi ^ {T} (s) \Gamma_ {g} ^ {T(l)} T_ {g}^{(l)} \Gamma_{g}^{(l)} W\beta_ {j} = \lambda \phi ^ {T} (s) \beta_{j}.
      $$
      Observing that the previous expression is valid for all values of $s$, one gets
      $$
      (n _ {g} ^ {(l)})^{-1} \Gamma_ {g} ^ {T(l)} T_ {g}^{(l)} \Gamma_ {g}^{(l)} W\beta_{j} =\lambda \beta_ {j},
      $$
      with the additional constraint $ \Vert \psi_j \Vert ^ {2} = 1 $ that turns into $\beta^ {T} W\beta = 1 $. Let us define
      $ u_ {j} = W ^ {1/2} \beta_ {j}$ and, then, the following  eigenequation is finally
      obtained:
     $$
      (n _ {g} ^ {(l)})^{-1}  W ^ {1/2} \Gamma_ {g} ^ {T(l)} T_ {g}^{(l)} \Gamma_ {g}^{(l)}W ^{1/2} u_ {j} = \lambda u_ {j}
     $$
     subject to $ u ^ {T} u_ {j} = \|u_j\|^2=1 $. From this equation we can compute the eigenvalues $\lambda_{j}$ and the
     vector of coefficients $ \beta_{j} = W ^{-1/2} u_{j}$, with which we
     calculate the eigenfunctions $ \psi_{j} (s) $ and the principal
     component scores $ C_{jg}^{(l+1)} $ are given by $C_{jg}^{(l+1)} = \Gamma_{g}^{(l)} W \beta_{j}$.

      \item[2.2.3] \textit{Scatter parameters update:} The parameters $a_{1,g},...,a_{q_{g},g}$ and $b_{g}$ are initially
      estimated as $\hat{a}_{j,g}^{(l+1)}=\lambda_{j}$ for the $q_g$ first eigenvalues of the
      $W^{1/2}\Gamma_{g}^{T(l)}T_{g}^{(l)}\Gamma_{g}^{(l)} W^{1/2}$ matrix and
      $$
      b_g^{(l+1)} =\frac{1}{p-q_{g}}\left[\text{trace}\left(W^{1/2}\Gamma_{g}^{T(l)}T_{g}^{(l)}\Gamma_{g}^{(l)}W^{1/2}\right)-\sum_{j=1}^{q_{g}} \hat{a}_{jg}^{(l+1)}\right].
      $$
      Recall that the scatter parameters so obtained do not necessarily satisfy the required constraints for
      the given $d_1$ and $d_2$ constants. In case that these constraints do not hold,  following \cite{Fritz2013},
      we define their truncated versions as:
    $$
     a_{jg}^{m_{1}}=
      \begin{cases}
       a_{jg} &\mbox{if  } a_{jg}\in [m_{1},d_{1}m_{1}],\\
       m_{1} &\mbox{if  } a_{jg} < m_{1},\\
       d_{1}m_{1} &\mbox{if  } a_{jg} > d_{1}m_{1},\\
     \end{cases}
    $$
    and
    $$
     b_{g}^{m_{2}}=
      \begin{cases}
      b_{g} &\mbox{if  } b_{g}\in [m_{2},d_{2}m_{2}],\\
      m_{2} &\mbox{if  } b_{g} < m_{2},\\
      d_{2}m_{2} &\mbox{if  } b_{g} > d_{2}m_{2}.\\
    \end{cases}
   $$
   The scatter parameters are finally updated as
   $\{a_{1,g}^{m_{opt_{1}}},...,a_{q_{g},g}^{m_{opt_{1}}},b_{g}^{m_{opt_{2}}},...,b_{g}^{m_{opt_{2}}}\}$
   where $m_{opt_{1}}$ minimizes
   $$
   m_{1} \mapsto
   \sum_{g=1}^{K}n_{g}\sum_{j=1}^{q_{g}}\left(\log(a_{jg}^{m_{1}})+\frac{a_{jg}}{a_{jg}^{m_{1}}}\right),
   $$
   and, $m_{opt_{2}}$ minimizes
   $$
   m_{2} \mapsto \sum_{g=1}^{K}n_{g}(p-q_{g})\left(\log(b_{g}^{m_{2}})+\frac{b_{g}}{b_{g}^{m_{2}}}\right)
   $$
   where $n_{g}=\sum_{i=1}^{n}\tau_{ig}$. 
   These are indeed two real-valued functions that can be easily minimized (see \cite{Fritz2013}).
   \end{itemize}
   \end{itemize}

  \item[3.] \textit{Evaluate target function:}
  After applying the trimmed EM steps, the associated value of the target function \eqref{eq11}
  is computed (we set $\eta(x_i)=0$ if $i\in I$ and $\eta(x_i)=1$ if
  $i\notin I$ for $I$ defined as in (\ref{i}) with the final iteration parameters).
  The set of parameters yielding
  the highest value of this target function and the associated trimmed
  indicator function $\eta$ are returned as the final algorithm's output.
\end{enumerate}

\subsection{Estimation of dimensions}\label{BIC_section}
As in \cite{BJ2011} and \cite{JP2013}, the estimation of the
dimensions per group, $q_ {g}$, $g = 1,...,K $ in the K-L expansion
is not an easy task and still an open problem. In the previously
mentioned works, the authors used the ``Cattell" procedure
\citep{Catell1966} and show that, by using an appropriate threshold,
$K$ sensible values can be obtained. However, the application of the
``Cattell" threshold within the EM algorithm may create increments
and decrements of the target function between two successive
iterations. In this work, we prefer solving the maximization of the
target function for fixed combinations of dimension and, later,
choose the dimensions yielding the better value of a penalized
likelihood for fixed values of trimming levels $\alpha $ and
constraints $d_{1}$ and $d_{2}$. To be more precise, we choose the
dimensions minimizing the Bayesian Information Criterion (BIC)
defined as
$$
 BIC = -2 l ^ {p}_{\alpha} (\hat {\theta};x_1,...,x_n) + \kappa \log (n)
$$
where $ l ^ {p}_{\alpha}  (\hat {\theta}) $ corresponds to the
trimmed log-likelihood function valued at the estimated optimal
parameters $\hat{\theta}$, $n$ is the number of observations and
$\kappa$ corresponds to the number of free parameters to be
estimated. We have $ \kappa = \rho + \nu + 2K + Q $, where $\rho =
(Kp + K-1)$ is the number of parameters needed to estimate means and
mixture proportions, $\nu = \sum_ {g = 1} ^ {K} q_ {g} [p- (q_ {g}
+1) / 2]$ corresponds to the number of parameters needed to estimate
the $ \psi_{j}$ eigenfunctions and $Q = \sum_ {g = 1} ^ {K} q_ {g}$.

\begin{figure}[htp]
    \centering
    \includegraphics[scale=.45]{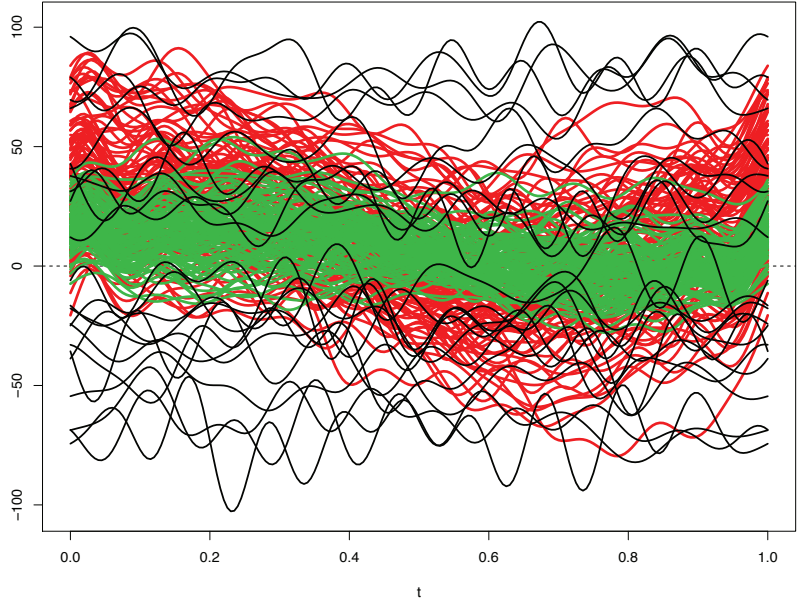}
    \caption{A simulated data set with $K=2$ groups from ``Scenario 1 and contamination scheme (iii)" (as described in Section \ref{Simustudy}).
    The subspace dimensions in this example are $q_1=2$ and $q_2=3$.}
    \label{fig:dataSBIC}
\end{figure}

\begin{figure}[htp]
\centering
\begin{tabular}{cc}
\includegraphics[scale=.35]{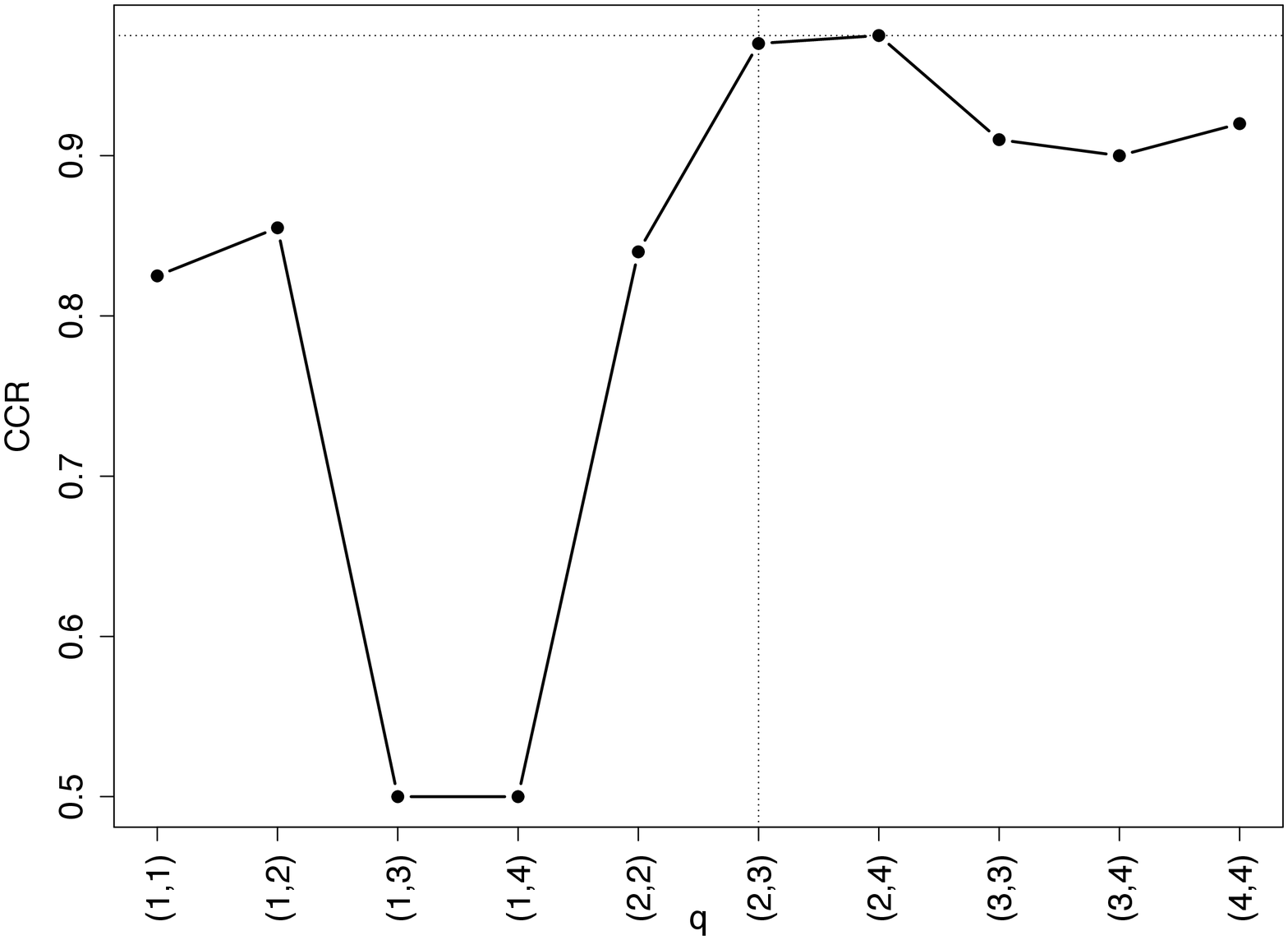}& \includegraphics[scale=.35]{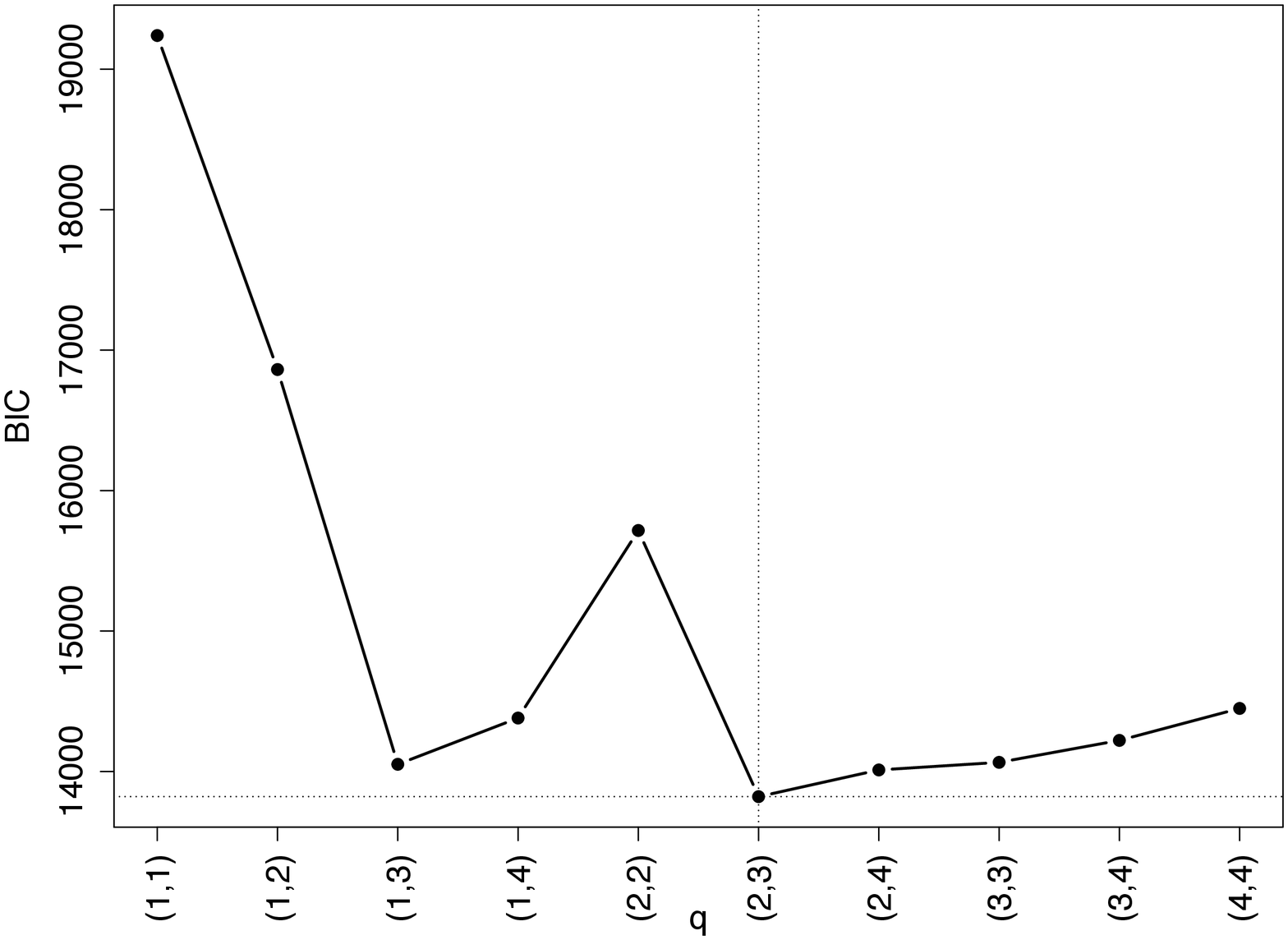} \\
\end{tabular}
\caption{Selection of the dimensions by means of BIC for the data set in Figure 1 when $d_{1}=d_{2}=10$ and $\alpha=0.1$. Different  combinations of dimensions are represented in the x-axis,  BIC values are presented in the right panel while the corresponding correct classification rates are in the left panel.
}
\label{fig:dataCCRBIC}
\end{figure}

To illustrate the use of BIC in the selection of the dimensions $q$
in the K-L expansion, we simulate a data set from the simulation
scheme called Scenario 1  ($q_1=2$ and $q_2=3$) with $10\%$
contamination of type iii) as will be fully described in the Section
\ref{Simustudy}). Figure \ref{fig:dataSBIC} shows that simulated
dataset. Figure \ref{fig:dataCCRBIC} (right panel) shows the BIC
values for several dimension combinations when $d_{1}=d_{2}=10$ and
$\alpha=0.1$. After, testing those combinations, it was observed
that the minimum value for the BIC corresponds to dimension $q_1=2$
and $q_2=3$. Moreover, as can be seen in Figure \ref{fig:dataCCRBIC}
(left panel), we note that the minimum value of the BIC corresponds
to one of the best solutions of the algorithm in terms of correct
classification rate (CCR).

\section{Simulation study}\label{Simustudy}
In order to evaluate the performance of the methodology proposed, we
simulated different scenarios and contamination types.

For the ``good" observations arranged in $K=2$ clusters, we consider
the following scheme of simulation based on the K-L expansion:
 \begin{align}\label{sim}
 x_i(t)=\mu_{g}(t)+\sum_{j=1}^{q_{g}}a_{jg}^{1/2}z_i\psi_{j}(t)+\sum_{j=q_{g}+1}^{p}b_{g}^{1/2}z_i\psi_{j}(t)\;\;\; t\in [0,1],
\end{align}
where $z_i$ are independent and $N(0,1)$-distributed, $\mu_{g}$ are
the group mean function, $a_{jg}$ corresponds to the main variances
and $b_{g}$ corresponds to the residual variability. In this
simulation, we consider that the eigenfunctions $\psi_{j}$ are the
first 21 Fourier basis functions that are defined as
\begin{align}\nonumber
\psi_{j}(t)=
\begin{cases}
\psi_{0}(t)=1,\\
\psi_{2j-1}(t)=\sqrt{2}\sin(j2\pi t)\\
\psi_{2j}(t)=\sqrt{2}\cos(j2\pi t)
\end{cases},
\end{align}
for $j=1,2,...,p$. We assume that the first $i=1,...,100$
observations are generated when $g=1$ in \eqref{sim} and the second
group of observations with indices $i=101,...,200$ are generated when $g=2$ in
\eqref{sim}. We have two different main scenarios for the ``good"
part of data depending on the mean functions and chosen variances:

\begin{itemize}
 \item[]\emph{Scenario 1:}  The groups have the same mean $\mu_{1}(t)=\mu_{2}(t)=\cos(t)$ and dimensions $q_1=2$ and $q_2=3$.
 The variances for the first group are $(a_{11},a_{21})=(60,30)$ and $b_{1}=0.5$. For the second group, the variances are $(a_{12},a_{22},a_{32})=(170,140,120)$ and $b_{2}=1$.
 \item[]\emph{Scenario 2:}  The groups have different means $\mu_{1}(t)=\cos(t)+3$ and $\mu_{2}(t)=\cos(t)+1$ and
 the dimensions are $q_1=2$ and $q_2=3$. The variances are $(a_{11},a_{21})=(a_{12},a_{22})=(60,30)$, $b_{1}=0.5$ and $b_{2}=1$.
\end{itemize}

We also consider the possibility of adding another 22 curves ($10\%$
contamination level) to see the effect of noise in clustering. In
two out of the three contaminating schemes, each of these 22
contaminating curves $x_i$ are obtained by fitting a linear
combination of the 21 first Fourier base elements plus a global mean
which interpolates 21 points in $\mathbb{R}^2$ as
$$
\{(t_l,u_i+\varepsilon_{i,l})\}_{l=1}^{21},
$$
where $\{t_l\}_{l=1}^{21}$ is an equispaced grid on $[0,1]$,
$\{u_i\}_{i=1}^{22}$ is the result of random sample from a uniform
distribution in the $[a,b]$ interval (to be specified latter) and
$\{\varepsilon_{i,l}\}_{l=1}^{21}$, for $i=201,...,222$, are
independent normally distributed error terms with variance
$\sigma^{2}=10$.

For both scenarios (Scenario 1 and 2) for the ``good" part of data,
we consider the following contaminating schemes:
\begin{itemize}
\item[(i)] No contamination (i.e., the total number of observations is $n=200$).
\item[(ii)] Using the previously described contaminating scheme with
$[a,b]=[150,180]$. This means that the contaminating curves are
clearly far apart from the ``good" curves.
\item[(iii)] Using the previously described contaminating scheme with
$$[a,b]=\left[\min_{i=1,...,200;t\in[0,1]}x_i(t),\max_{i=1,...,200;t\in[0,1]}x_i(t)\right].$$
\item[(iv)] We use the scheme in \eqref{sim} also for $i=201,...,222$
but the normally distributed $z_{ij}$ variables are replaced by
(heavier tailed) Cauchy distributed ones.
\end{itemize}

In order to test the performance of the methodology proposed here,
we carry out a simulation study using the scheme previously
described and compare the results with those obtained by ``Funclust"
\citep{JP2013} and ``FunHDDC" \citep{BJ2011}.

In this simulation study, it is important to note that we assume the
$q_g$ dimensions to be unknown parameters and that we use the BIC
proposal described in section \ref{BIC_section} to estimate them when
applying the proposed robust functional clustering (RFC). We use
trimming levels $\alpha=0$ (untrimmed) and $\alpha=0.1$, constraints
$d_{1}=d_{2}=1$, $d_{1}=d_{2}=10$ and $d_{1}=d_{2}=10^{10}$ (i.e.,
almost unconstrained in this last case). We always return the best
solution in terms of the highest BIC value for each combination of
all those fixed values of trimming levels and constraints. We use
\texttt{niter}$=100$ random initializations with \texttt{iter.max}$=20$.

For the ``Funclust" method we have used the library
\texttt{Funclustering} \citep{funclustR} in R where the EM algorithm
has been initialized with the best solutions out of 20 ``short" EM
algorithms with only 20 iterations with values of
$\varepsilon=0.001, 0.05, 0.1$ in the Cattell test. In the case of
the ``FunHDDC", we use the library \texttt{FunHDDC} \citep{funHDDCR}
in R with values of $\varepsilon=0.001, 0.05, 0.1$ in the Cattell
test, moreover, the submodels $A_{kj}B_{k}Q_{k}D_{k}$,
$A_{k}B_{k}Q_{k}D_{k}$, $A_{k}BQ_{k}D_{k}$, $ AB_{k}Q_{k}D_{k}$,
$ABQ_{k}D_{k}$ are tested, see details in \citep{BJ2011} and the
best solution in terms of the highest BIC value for all those
submodels are returned.

\begin{figure}[htp]
   \centering
   \includegraphics[scale=.65]{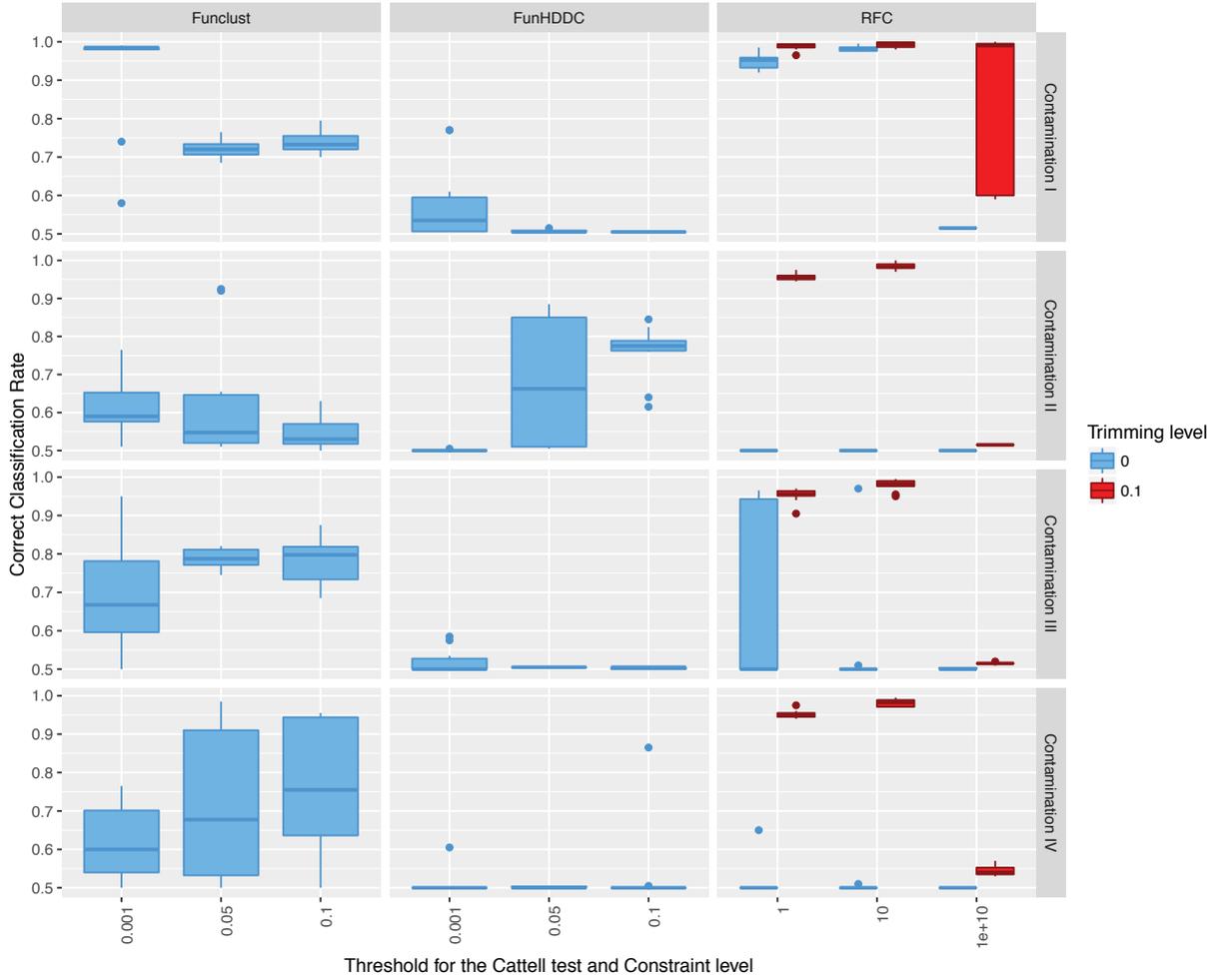}
   \caption{Scenario 1 (equal mean functions): Correct classification rate (CCR) for the three methods considered, represented in different columns. Rows correspond to the different contamination schemes (i) to (iv), described previously in this section, starting with no contamination in the first row. Constraint levels $d_1=d_2=1$, $10$ and $10^{10}$ and trimming levels $\alpha=0$ and $0.1$ are used for the RFC method and the proposed BIC to choose dimension. Threshold values $\varepsilon= 0.001, 0.05$ and $0.1$ are used for the ``Cattell" procedure in ``Funclust" and ``FunHDDC".
}
   \label{fig:SSMI}
\end{figure}

Figure \ref{fig:SSMI} shows the results for datasets simulated
according to Scenario 1, i.e. groups with equal means. This figure
is composed of a matrix of graphs, where the rows correspond to the
different contamination schemes (uncontaminated in the first row)
while the columns correspond to the methodologies tested. The first
column corresponds to ``Funclust", the second to ``FunHDCC" and the
third one shows the results for the robust functional clustering
(RFC) procedure with the two different trimming levels and the three
constraints levels (we are assuming $d_1=d_2$ to simplify the
simulation study). The $x$-axis corresponds to the threshold in the
Cattell test for the first two columns, and to the constraint level
for RFC, while the $y$-axis corresponds to the correct
classification rate (CCR).

The results show that the joint use of trimming and constraints in
RFC improve the CCR substantially. Results are very good for
moderate $(d_{1}=d_{2}=10)$ and small $(d_{1}=d_{2}=1)$ values of
the constraint constants, while for high values the results are
poor. Very high values for these constants are equivalent to having
unconstrained parameters. The use of trimming also turns out to be
very useful in all the contaminated case while it does not affect so
much the results in the uncontaminated case.

In most cases the results for ``FunHDDC" and ``Funclust" fall below
those of RFC when applying the $\alpha=0.1$ trimming and
small/moderate values $d_1$ and $d_2$ for the variance parameters.
The only case where this is not so is ``Funclust" with $\tau=0.001$
in the first row, corresponding to uncontaminated data. However,
this method requires the use of $q_{1}=20$ and $q_{2}=8$ terms in
the K-L expansion for groups 1 and 2 respectively.

\begin{figure}[htp]
    \centering
    \includegraphics[scale=.65]{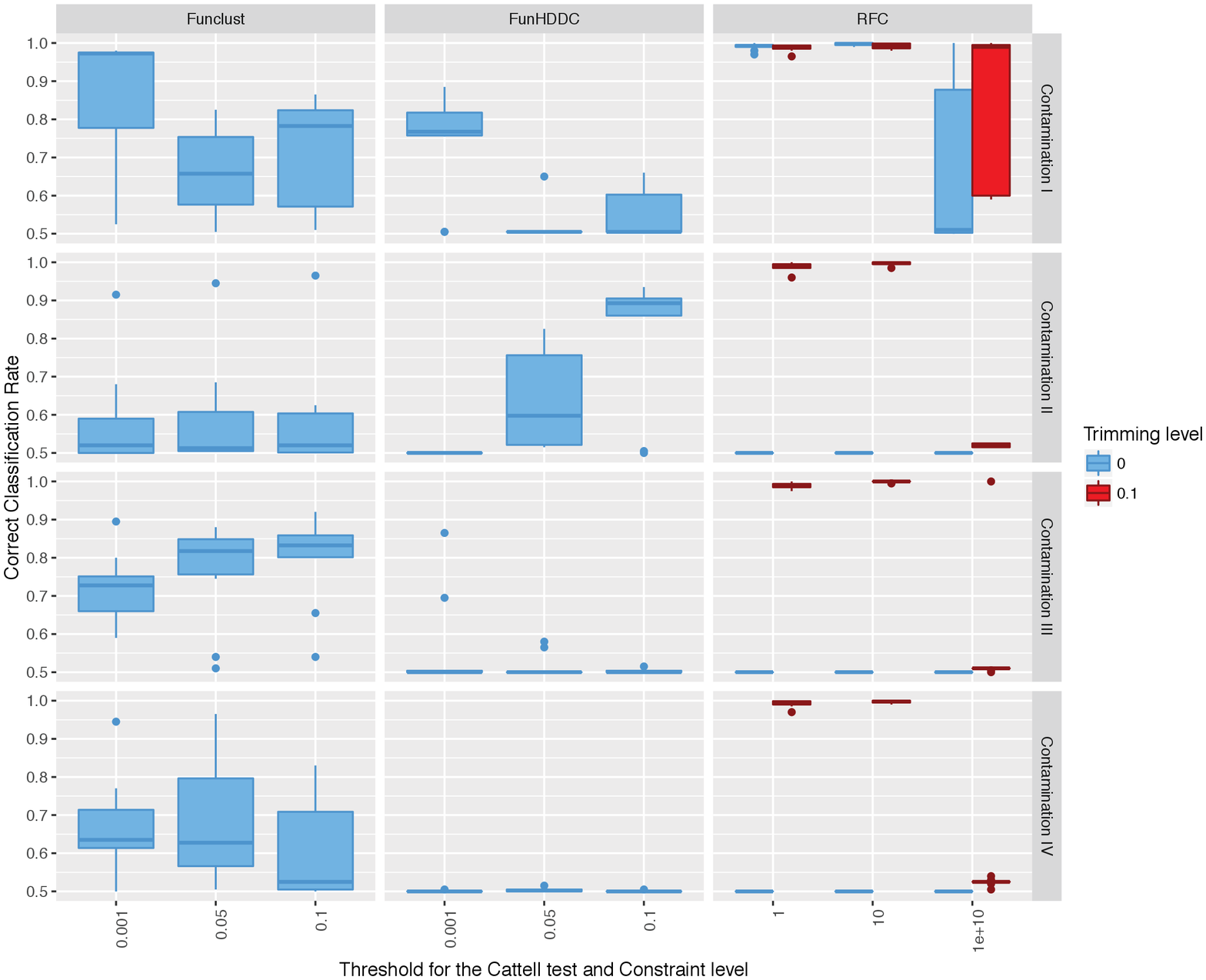}
        \caption{Scenario 2 (unequal mean functions): Correct classification rate (CCR) for the three methods considered, represented in different columns. Rows correspond to the different contamination schemes (i) to (iv), described previously in this section, starting with no contamination in the first row. Constraint levels $d_1=d_2=1$, $10$ and $10^{10}$ and trimming levels $\alpha=0$ and $0.1$ are used for the RFC method and the proposed BIC to choose dimension. Threshold values $\varepsilon= 0.001, 0.05$ and $0.1$ are used for the ``Cattell" procedure in ``Funclust" and ``FunHDDC".}
    \label{fig:SSSMD}
\end{figure}

The results corresponding to Scenario 2 are presented in Figure
\ref{fig:SSSMD}. This scenario corresponds to groups with different
means. Again, it can be seen that the joint use of trimming and
constraints improve the results in terms of classification rates.
The results in these cases, both for moderate $(d_{1}=d_{2}=10)$ and
small $(d_{1}=d_{2}=1)$ values of the constraint constants are quite
good, while the results are poor for very large $d_1=d_2$ values. In
this case the RFC method with appropriate trimming and constraints
always performs better than ``FunHDDC" and ``Funclust" in terms of
classification accuracy.\\

In addition, it is worth mentioning that the results for RFC method for both simulation scenarios are more consistent, in the sense that the correct classification rate (CCR) has a lower dispersion for this method, which indicates another advantage of this proposal for robust clustering.

\section{Real data example: NOx levels}\label{real_data}
The data set corresponds to daily curves of Nitrogen Oxides NOx
emissions in the neighborhood of the industrial area of Poblenou,
Barcelona (Spain). NOx is one of the principal contaminant agents
and characterizing its behavior is useful to develop appropriate
environmental policies. The detection of outlying emission curves
from any data source is meaningful because the explanation of why
these curves are observed may be helpful in order to forecast or
anticipate them. In addition, these outlying curves can also
influence non-robust clustering methods leading to wrong conclusions
when searching for clusters of days with different types of emission patterns.

The data are available in the \texttt{fda.usc} library
\citep{FBfdausc} in R. The measurements of NOx (in $\mu g/m^3$) were
taken hourly resulting in 115 days with complete observations. This
data set has been analyzed to test methodologies for the detection
of outliers in functional data in \cite{Foutnox2008},
\cite{Sguera2015} and \cite{RFPCA}.

\begin{figure}[htp]
    \centering
    \includegraphics[scale=.5]{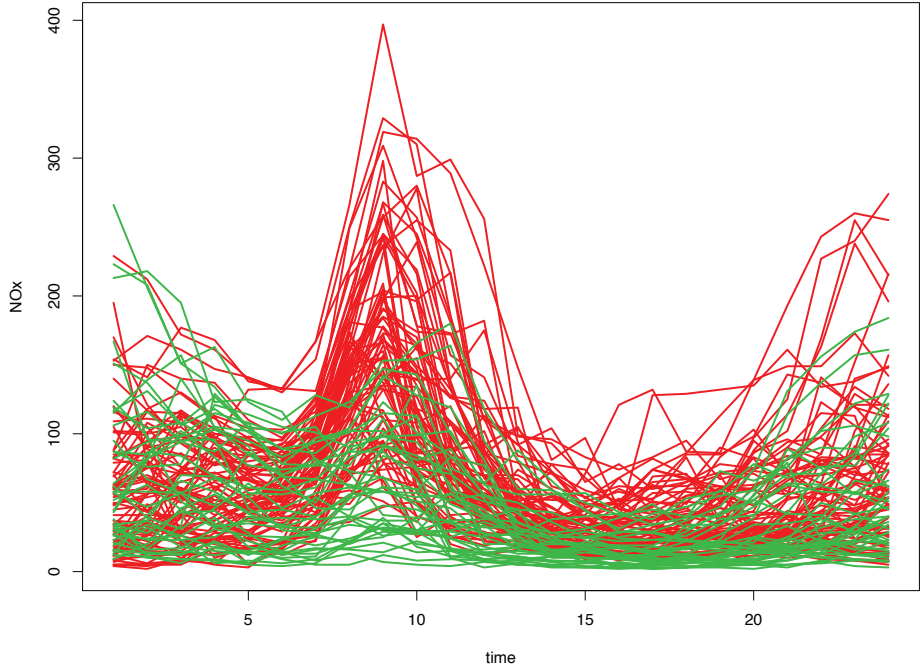}
        \caption{Curves represent daily levels of NOx for 115 days, with 76 working day in red and 39 non-working days in green.}
    \label{fig:nox}
\end{figure}

Figure \ref{fig:nox} shows the original daily curves of Nitrogen
Oxides NOx emissions by using different colors, red for the 76
``working days" and green for the 39 ``non-working days".

The RFC methodology is applied to this dataset and the results are
compared to those obtained using the ``Funclust" and ``FunHDCC"
methodologies. Two clusters ($K=2$) and a B-spline basis of functions
of order $3$ with $15$ basis elements (13 equispaced knots) are
taken. For RFC, we use trimming levels $\alpha=0$, $0.1$ and $0.15$,
and constraints values $d_{1}=d_{2}$ equal to $1$, $10$ and
$10^{10}$ (\texttt{nstart}$=100$ and \texttt{niter}$=20$). For the
``Funclust" and ``FunHDCC" methods, we use the same strategy as in
the simulation study with values of $\varepsilon=0.001$, 0.05 and
0.1 in the Cattell procedure. The dimensions are estimated by using the
BIC criterion for the RFC method and also when applying the ``FunHDDC" method.

Table \ref{tab:resnox} shows a summary of the results obtained for
different combinations of input parameters. The second column of the
table shows the estimated $q_{g}$ dimensions by means of the BIC for
RFC method and the Cattell test for ``Funclust" and ``FunHDDC" methods.
We also give the correct classification rates (``CCR" column)
assuming that the ``true" clusters in data were only determined by
the type of day (working and non-working days). In this column,
again, we are re-assigning the trimmed observations according to
their posterior probabilities of membership to clusters.

\begin{table}[htp]
\centering
\begin{tabular}{lcrrrrcc}
\hline
           & $q_{g}$ &$\alpha$ &$d_{1}$&$d_{2}$&$\varepsilon$& CCR\\
\hline
       RFC &  2,5 &     0 & 1 & 1 &-&  0.84\\
           &  5,5 &     0 & 10 & 10 &-&0.70  \\
           &  5,5 &     0 & $10^{10}$ & $10^{10}$ &-& 0.69 \\\cline{2-7}
           &  2,5 &     0.1 & 1 & 1 &-&  0.85\\
           &  5,5 &     0.1 & 10 & 10 &-& 0.69 \\
           &  5,5 &     0.1& $10^{10}$ & $10^{10}$ &-& 0.66 \\\cline{2-7}
           &  2,5 &     0.15 & 1 & 1 &-&  0.84\\
           &  5,5 &     0.15 & 10 & 10 &-& 0.70\\
           &  5,5 &     0.15 & $10^{10}$ & $10^{10}$ &-& 0.69\\

\hline
Funclust  & 14,13 & -&- &-&0.001  &  0.84 \\
          & 4,5   & -&- &-&0.05 & 0.66 \\
          & 3,3   & -&- &-&0.1 &0.66          \\
\hline
FunHDDC  & 14,10 & -& - &-&0.001& 0.66    \\
          &3,2 & - &- &-&0.05 &  0.66  \\
          & 1,3 & - & -&-&0.1 &  0.66 \\
\hline

\end{tabular}
\caption{Correct classification rate (CCR) and dimension estimated for different levels of trimming $\alpha$ and constraints $d_{1}$ and $d_{2}$, for the RFC method and different vales of  $\varepsilon$ for the Cattell test in ``Funclust" and ``FunHDDC".}
 \label{tab:resnox}
\end{table}

One can see that the use of strong $d_1=d_2=1$ constraints slightly
increases the CCR (assuming that the correct groups were determined by
working and non-working days). In this case, the CCR for RFC is
84.3$\%$ without trimming and 85$\%$ with a $\alpha=0.1$ and
84.3$\%$ with $\alpha=0.15$ while the best CCR for ``Funclust" is
$84.3\%$. However, the RFC method has an additional advantage in
that it requires smaller dimensions than
``Funclust" for achieving that level of CCR.

Another important point is that the RFC allows us to perform
clustering and outlier detection simultaneously while ``Funclust"
and ``FunHDDC" do not. Even though the detected outliers are not so
extreme in this case as to completely deteriorate the clustering
process, it is also interesting to detect these outlying curves also
taking the cluster structure in mind. In this direction, every
trimmed curve (trimming levels $\alpha=0.1$ and $\alpha=0.15$)
corresponds to outliers already detected in previous works in the
literature that were also concerned with functional outlying
detection as \cite{Foutnox2008}(DEPTH), \cite{Sguera2015}(KFSD) and
\cite{RFPCA}(BACONPCA). Two separated data sets, considering only
working days (W) and non-working days (NW), were considered when
applying \cite{Sguera2015}(KFSD) while the complete dataset, without
differentiating between working and non-working days (W-NW), is used
when applying our RFC proposal and the other two methods.


 \begin{table}[htbp]
 \centering
 \begin{tabular}{cccccc}
 \hline
 \multicolumn{2}{c}{KFSD}   & \multicolumn{1}{c}{BACONPCA} & \multicolumn{1}{c}{DEPTH} & \multicolumn{1}{c}{RFC $\alpha=0.1$}&\multicolumn{1}{c}{RFC$\alpha=0.15$} \\ 
 NW & W & W-NW  & W-NW & W-NW & W-NW  \\ \hline
 12/03/2005 & 09/03/2005 & 18/03/2005 & 11/03/2005 & 25/02/2005&25/02/2005 \\ 
 19/03/2005 & 11/03/2005 & 29/04/2005 & 18/03/2005 & 03/03/2005&03/03/2005 \\ 
 30/04/2005 & 15/03/2005 & 11/03/2005 & 29/04/2005 & 11/03/2005&09/03/2005 \\ 
 01/05/2005 & 16/03/2005 & 02/05/2005 & 02/05/2005 & 16/03/2005&11/03/2005 \\ 
            & 17/03/2005 & 09/03/2005 &            & 18/03/2005&16/03/2005 \\ 
            & 18/03/2005 &            &            & 25/04/2005&18/03/2005 \\ 
            & 29/04/2005 &            &            & 29/04/2005&18/04/2005 \\ 
            & 02/05/2005 &            &            & 02/05/2005&25/04/2005 \\ 
            &            &            &            & 18/05/2005&29/04/2005 \\ 
            &            &            &            & 27/05/2005&02/05/2005 \\ 
            &            &            &            & 23/06/2005&03/05/2005 \\ 
            &            &            &            & 15/05/2005&18/05/2005 \\
            &          &            &            &           & 27/05/2005\\
            &            &            &            &           &23/06/2005 \\
       &            &            &            &           &19/03/2005 \\
       &            &            &            &           &30/04/2005 \\
       &            &            &            &           &15/05/2005\\ \hline
 \end{tabular}
 \caption{Outliers detected when using
  \cite{Foutnox2008}(DEPTH), \cite{Sguera2015}(KFSD), \cite{RFPCA}(BACONPCA) and the
  proposed RFC methodology with $\alpha=0.1$ and $\alpha=0.15$. Separated data sets
considering only working days (W) and non-working days (NW) were
used by \cite{Sguera2015}(KFSD) while the complete data
set (W-NW) were used for the other methods.}
 \label{tab:resnoxout}
 \end{table}

\begin{figure}[htp]
 \centering
   \begin{tabular}{ccc}
   \includegraphics[scale=.2]{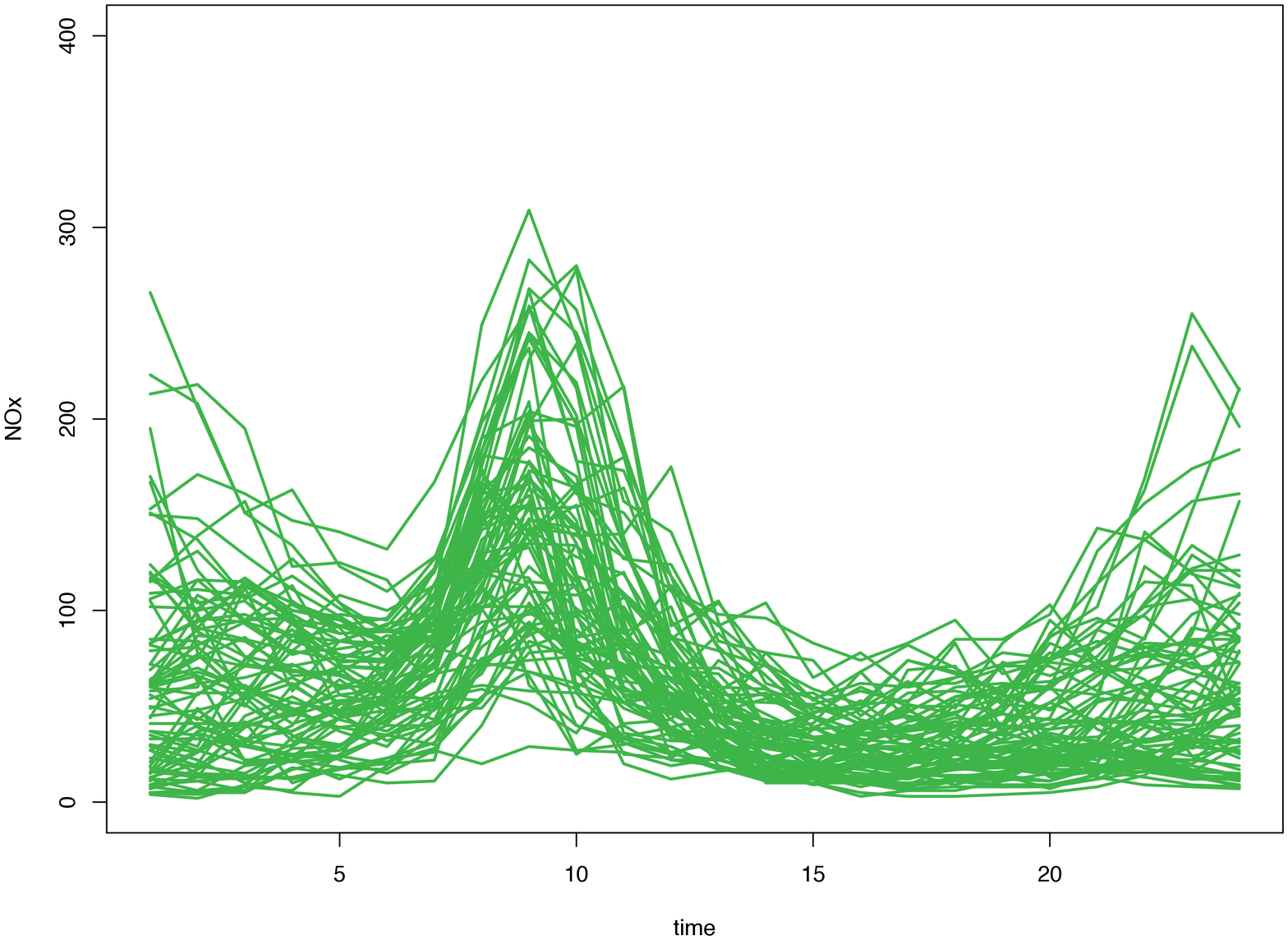}&\includegraphics[scale=.2]{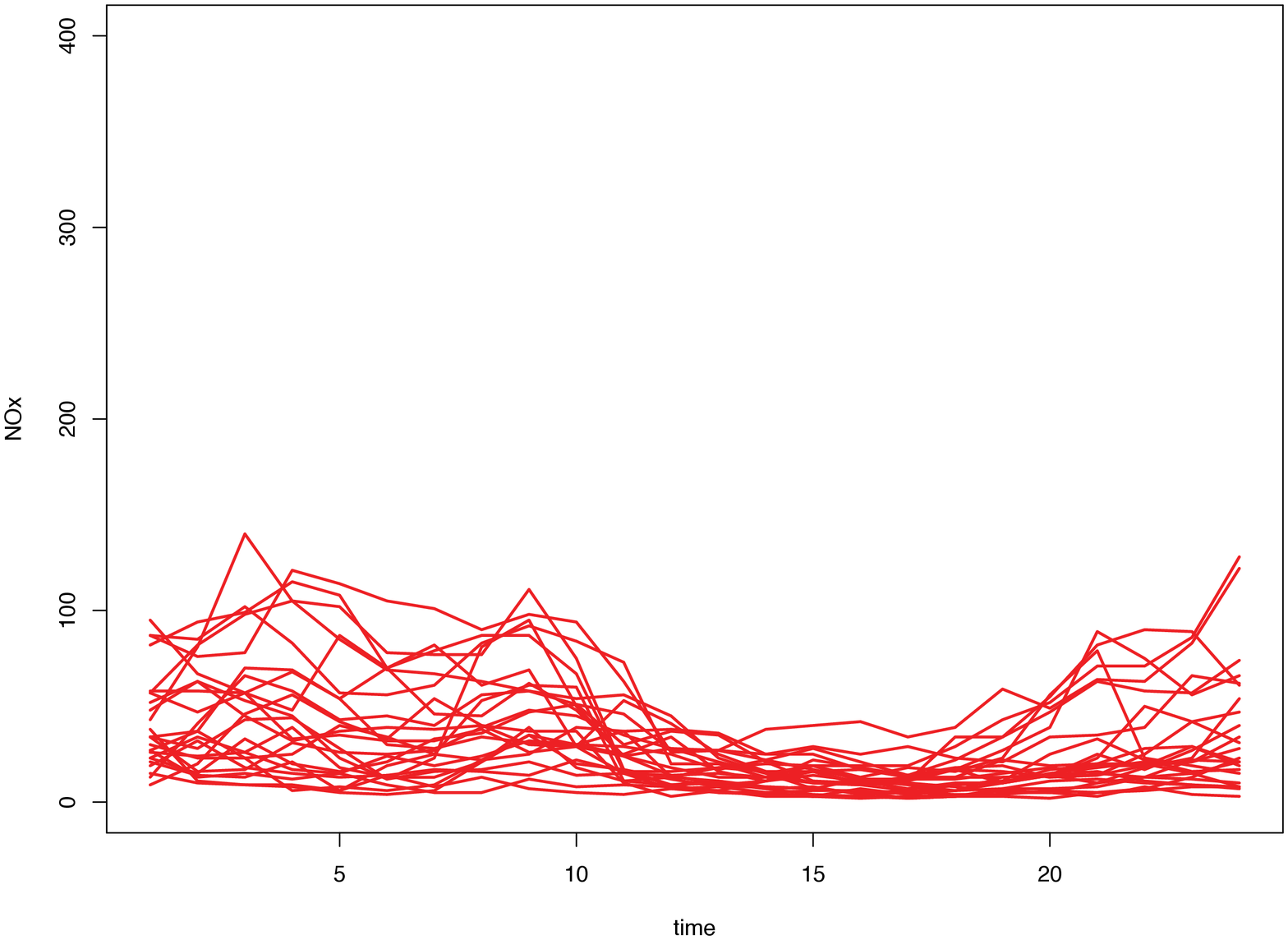}&\includegraphics[scale=.2]{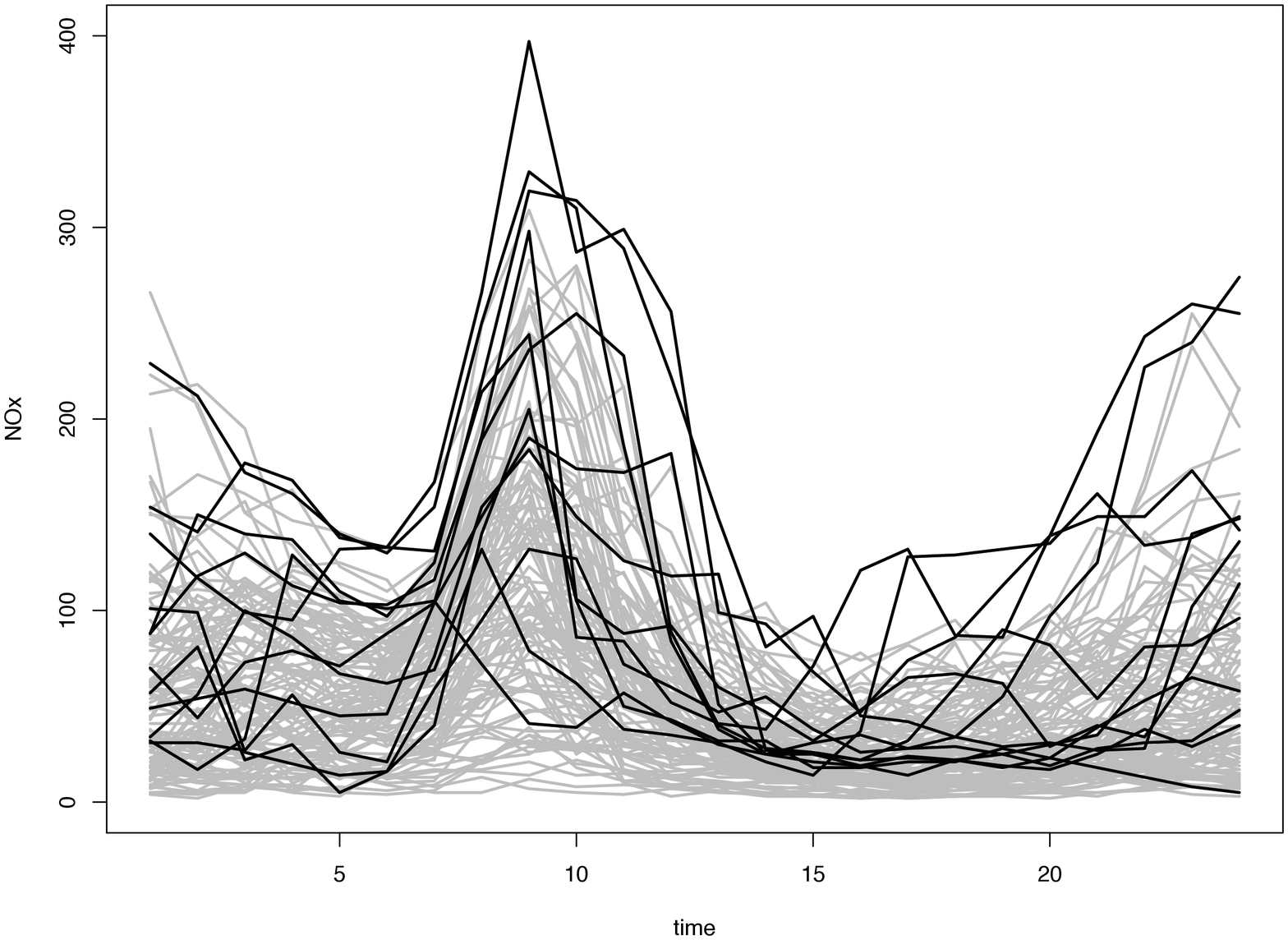}\\
   \includegraphics[scale=.2]{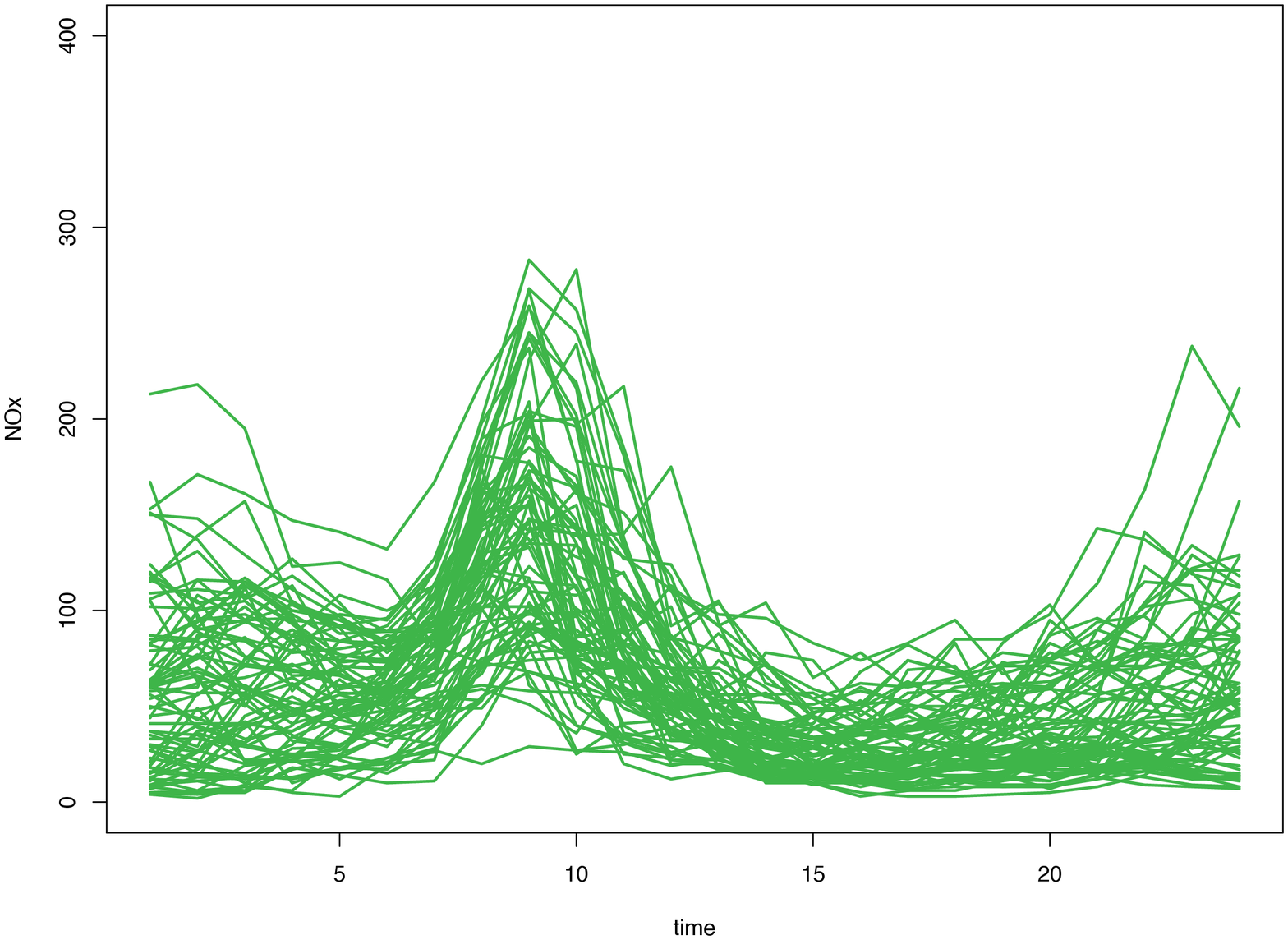}&\includegraphics[scale=.2]{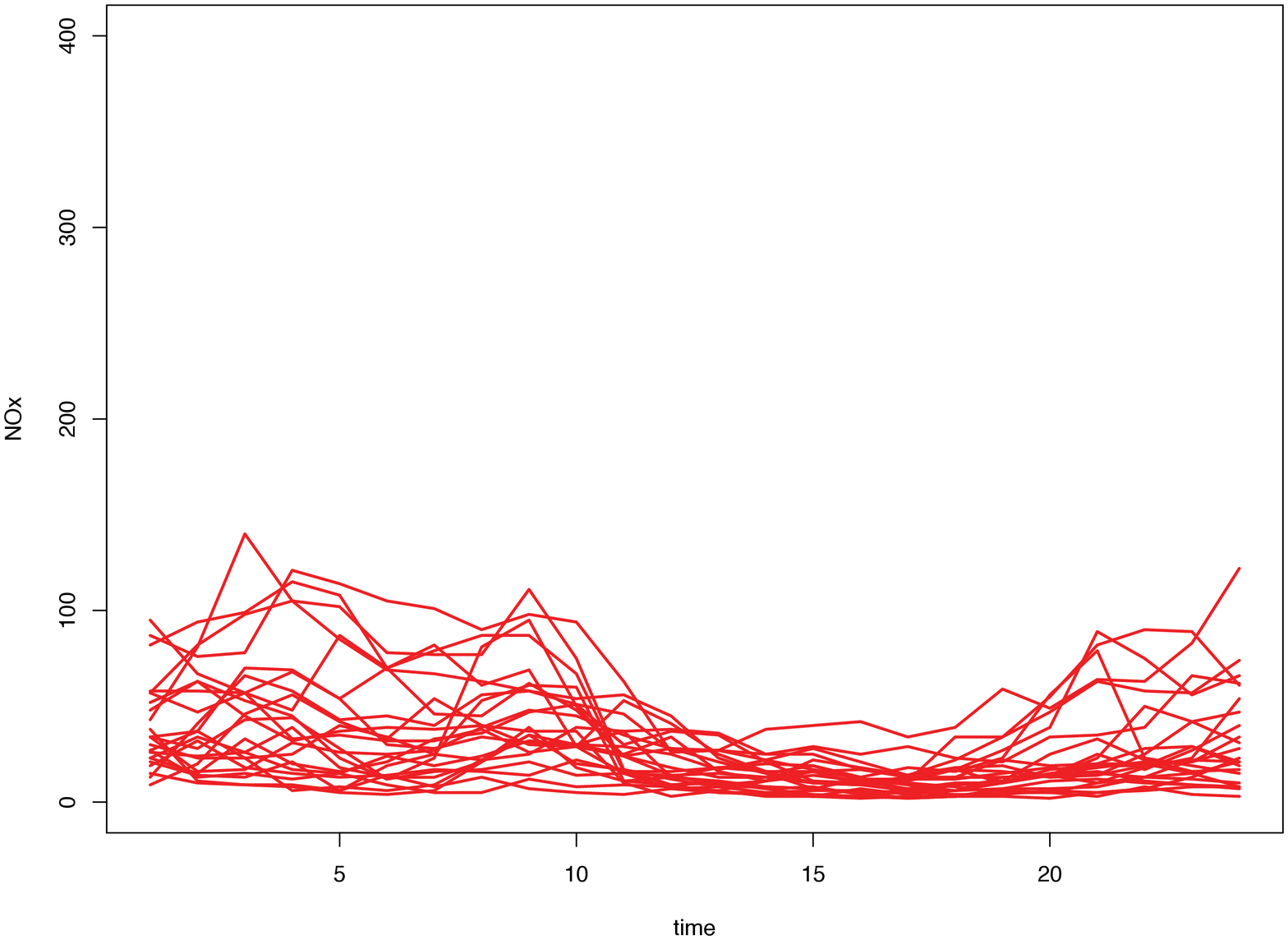}&\includegraphics[scale=.2]{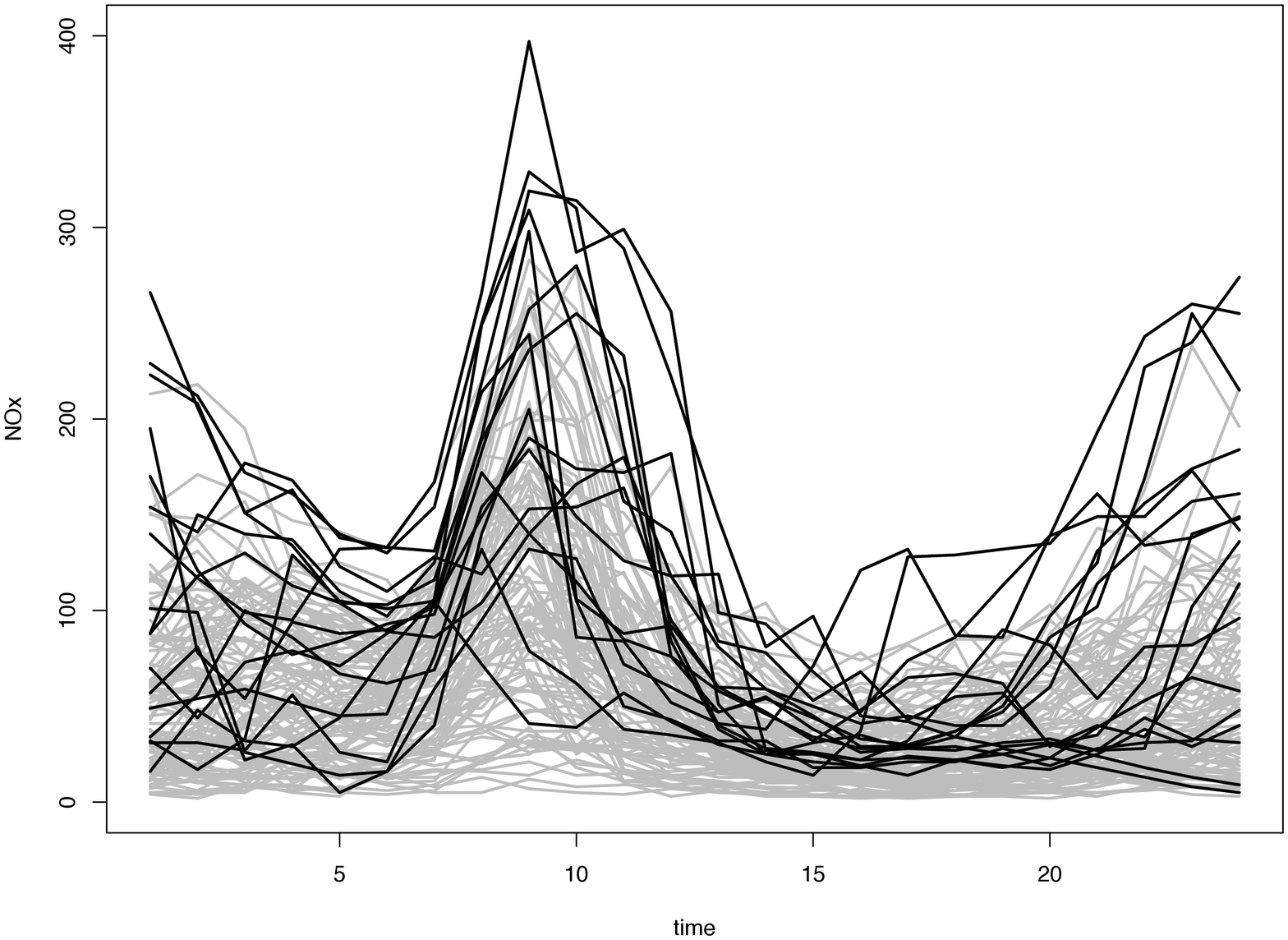}
   \end{tabular}
      \caption{Clusters found (non-trimmed curves in green and red) when applying the RFC method with $K=2$ and  $d_{1}=d_{2}=1$. The trimmed curves appear in
      black while the non-trimmed ones in gray. \textit{Top panels:} Trimming level $\alpha=0.1$.
      \textit{Bottom panels:} Trimming level $\alpha=0.15$.}
      \label{fig:resnox}
\end{figure}

Figure \ref{fig:resnox} shows the RFC clustering results. We observe
that the curves that are detected as outliers (in black in the third
column) exhibit different patterns from the rest of the
curves.
%
%

\section{Conclusions}\label{conclusions}
A feasible methodology for robust model-based functional clustering
has been proposed and illustrated. The key idea behind the algorithm
presented is the use of an approximation of the ``density" for
functional data together with the simultaneous use of trimming and
constraints. This allows for a robust model-based clustering
approach.

The use of trimming tools protects the estimation of the parameters
against the harmful effect of (even a small amount of) outlying
curves, while the constraints avoid the detection of spurious clusters
and improve the algorithm's stability. The simulation study shows
that the joint use of constraints and trimming tools improve the
performance of the clustering algorithm in comparison to some other
procedures for clustering functional data. The real data
example shows that the trimmed curves often correspond to
outliers already detected by other specialized methods for outlier
detection in functional data analysis. In fact, we conclude that the
proposed robust methodology can be  a useful
tool to detected contamination and groups in a functional data set
simultaneously.

However, some limitations of this methodology are the choice of level
of trimming $\alpha$ and the choice of the scatter constraints
constants $d_1$ and $d_2$. These values are subjective and sometimes
depend on the final purpose of the cluster analysis. For this
reason, we always recommend the use of different values of trimming
and constraint and monitoring the effect in the clustering partition
 of these choices. The development of more automatized
selection procedures for these values may be considered as an open problem for
future research.

Finally, an extension of our proposal for future work is the consideration of
multivariate functional data. 

\section{Acknowledgements}
This work was partly done while DR and JO visited the Departamento
de Estad\'{\i}stica e I.O., Universidad de Valladolid, Spain, with
support from Conacyt, Mexico (DR as visiting graduate student, JO by
Projects 169175 An\'alisis Estad\'{\i}stico de Olas Marinas, Fase II
y 234057 An\'alisis Espectral, Datos Funcionales y Aplicaciones),
CIMAT, A.C. and the Universidad de Valladolid. Their hospitality and
support is gratefully acknowledged. Research by LA G-E and A M-I was
partially supported by the Spanish Ministerio de Econom\'{\i}a y
Competitividad y fondos FEDER, grant  MTM2014-56235-C2-1-P, and  by
Consejer\'{\i}a de Educaci\'on de la Junta de Castilla y Le\'on,
grant VA212U13

\bibliography{references}

\begin{thebibliography}{}

\bibitem[\protect\citeauthoryear{Bouveyron and Jacques}{Bouveyron and
  Jacques}{2011}]{BJ2011}
Bouveyron, C. and J.~Jacques (2011).
\newblock Model-based clustering of time series in group-specific functional
  subspaces.
\newblock {\em Adv. Data Anal. Classif.\/}~{\em 5\/}(4), 281--300.

\bibitem[\protect\citeauthoryear{Bouveyron and Jacques}{Bouveyron and
  Jacques}{2014}]{funHDDCR}
Bouveyron, C. and J.~Jacques (2014).
\newblock {\em funHDDC: Model-based clustering in group-specific functional
  subspaces}.
\newblock R package version 1.0.

\bibitem[\protect\citeauthoryear{Cattell}{Cattell}{1966}]{Catell1966}
Cattell, R.~B. (1966).
\newblock The scree test for the number of factors.
\newblock {\em Multivariate Behaviour Research\/}~(2), 245--276.

\bibitem[\protect\citeauthoryear{Cuesta-Albertos and Fraiman}{Cuesta-Albertos
  and Fraiman}{2007}]{cuestafraiman2007}
Cuesta-Albertos, J.~A. and R.~Fraiman (2007).
\newblock Impartial trimmed {$k$}-means for functional data.
\newblock {\em Comput. Statist. Data Anal.\/}~{\em 51\/}(10), 4864--4877.

\bibitem[\protect\citeauthoryear{Cuesta-Albertos, Gordaliza, and
  Matr{\'a}n}{Cuesta-Albertos et~al.}{1997}]{CGmayo1997}
Cuesta-Albertos, J.~A., A.~Gordaliza, and C.~Matr{\'a}n (1997).
\newblock Trimmed {$k$}-means: an attempt to robustify quantizers.
\newblock {\em Ann. Statist.\/}~{\em 25\/}(2), 553--576.

\bibitem[\protect\citeauthoryear{Delaigle and Hall}{Delaigle and
  Hall}{2010}]{DH2010}
Delaigle, A. and P.~Hall (2010).
\newblock Defining probability density for a distribution of random functions.
\newblock {\em Ann. Statist.\/}~{\em 38\/}(2), 1171--1193.

\bibitem[\protect\citeauthoryear{Febrero, Galeano, and
  Gonz{\'a}lez-Manteiga}{Febrero et~al.}{2008}]{Foutnox2008}
Febrero, M., P.~Galeano, and W.~Gonz{\'a}lez-Manteiga (2008).
\newblock Outlier detection in functional data by depth measures, with
  application to identify abnormal {${\rm NO}_x$} levels.
\newblock {\em Environmetrics\/}~{\em 19\/}(4), 331--345.

\bibitem[\protect\citeauthoryear{Febrero-Bande and {Oviedo de la
  Fuente}}{Febrero-Bande and {Oviedo de la Fuente}}{2012}]{FBfdausc}
Febrero-Bande, M. and M.~{Oviedo de la Fuente} (2012).
\newblock Statistical computing in functional data analysis: The {R} package
  {fda.usc}.
\newblock {\em Journal of Statistical Software\/}~{\em 51\/}(4), 1--28.

\bibitem[\protect\citeauthoryear{Ferraty and Vieu}{Ferraty and
  Vieu}{2006}]{FV2006}
Ferraty, F. and P.~Vieu (2006).
\newblock {\em Nonparametric functional data analysis}.
\newblock Springer Series in Statistics. Springer, New York.

\bibitem[\protect\citeauthoryear{Fraley and Raftery}{Fraley and
  Raftery}{2002}]{Fraley2002}
Fraley, C. and A.~E. Raftery (2002).
\newblock Model-based clustering, discriminant analysis, and density
  estimation.
\newblock {\em J. Amer. Statist. Assoc.\/}~{\em 97\/}(458), 611--631.

\bibitem[\protect\citeauthoryear{Fritz, Garc{\'{\i}}a-Escudero, and
  Mayo-Iscar}{Fritz et~al.}{2013}]{Fritz2013}
Fritz, H., L.~A. Garc{\'{\i}}a-Escudero, and A.~Mayo-Iscar (2013).
\newblock A fast algorithm for robust constrained clustering.
\newblock {\em Comput. Statist. Data Anal.\/}~{\em 61}, 124--136.

\bibitem[\protect\citeauthoryear{Gallegos}{Gallegos}{2002}]{gallegos2002}
Gallegos, M.~T. (2002).
\newblock Maximum likelihood clustering with outliers.
\newblock In {\em Classification, clustering, and data analysis ({C}racow,
  2002)}, Stud. Classification Data Anal. Knowledge Organ., pp.\  247--255.
  Springer, Berlin.

\bibitem[\protect\citeauthoryear{Garc{\'{\i}}a-Escudero and
  Gordaliza}{Garc{\'{\i}}a-Escudero and Gordaliza}{2005}]{Efuncional2005}
Garc{\'{\i}}a-Escudero, L.~A. and A.~Gordaliza (2005).
\newblock A proposal for robust curve clustering.
\newblock {\em J. Classification\/}~{\em 22\/}(2), 185--201.

\bibitem[\protect\citeauthoryear{Garc{\'{\i}}a-Escudero, Gordaliza, Matr{\'a}n,
  and Mayo-Iscar}{Garc{\'{\i}}a-Escudero et~al.}{2008}]{GEV2008}
Garc{\'{\i}}a-Escudero, L.~A., A.~Gordaliza, C.~Matr{\'a}n, and A.~Mayo-Iscar
  (2008).
\newblock A general trimming approach to robust cluster analysis.
\newblock {\em Ann. Statist.\/}~{\em 36\/}(3), 1324--1345.

\bibitem[\protect\citeauthoryear{Garc{\'{\i}}a-Escudero, Gordaliza, Matr{\'a}n,
  and Mayo-Iscar}{Garc{\'{\i}}a-Escudero et~al.}{2015}]{Escudero2015}
Garc{\'{\i}}a-Escudero, L.~A., A.~Gordaliza, C.~Matr{\'a}n, and A.~Mayo-Iscar
  (2015).
\newblock Avoiding spurious local maximizers in mixture modeling.
\newblock {\em Stat. Comput.\/}~{\em 25\/}(3), 619--633.

\bibitem[\protect\citeauthoryear{Garc{\'{\i}}a-Escudero, Gordaliza, and
  Mayo-Iscar}{Garc{\'{\i}}a-Escudero et~al.}{2014}]{GEV2014}
Garc{\'{\i}}a-Escudero, L.~A., A.~Gordaliza, and A.~Mayo-Iscar (2014).
\newblock A constrained robust proposal for mixture modeling avoiding spurious
  solutions.
\newblock {\em Adv. Data Anal. Classif.\/}~{\em 8\/}(1), 27--43.

\bibitem[\protect\citeauthoryear{Jacques and Preda}{Jacques and
  Preda}{2013}]{JP2013}
Jacques, J. and C.~Preda (2013).
\newblock Funclust: A curves clustering method using functional random
  variables density approximation.
\newblock {\em Neurocomputing\/}~{\em 112}, 164--171.

\bibitem[\protect\citeauthoryear{James and Sugar}{James and
  Sugar}{2003}]{jamessugar2003}
James, G.~M. and C.~A. Sugar (2003).
\newblock Clustering for sparsely sampled functional data.
\newblock {\em J. Amer. Statist. Assoc.\/}~{\em 98\/}(462), 397--408.

\bibitem[\protect\citeauthoryear{Ramsay and Silverman}{Ramsay and
  Silverman}{2005}]{RS2006}
Ramsay, J.~O. and B.~W. Silverman (2005).
\newblock {\em Functional data analysis\/} (Second ed.).
\newblock Springer Series in Statistics. Springer, New York.

\bibitem[\protect\citeauthoryear{Ritter}{Ritter}{2015}]{RG2015}
Ritter, G. (2015).
\newblock {\em Robust cluster analysis and variable selection}, Volume 137 of
  {\em Monographs on Statistics and Applied Probability}.
\newblock CRC Press, Boca Raton, FL.

\bibitem[\protect\citeauthoryear{Sawant, Billor, and Shin}{Sawant
  et~al.}{2012}]{RFPCA}
Sawant, P., N.~Billor, and H.~Shin (2012).
\newblock Functional outlier detection with robust functional principal
  component analysis.
\newblock {\em Comput. Statist.\/}~{\em 27\/}(1), 83--102.

\bibitem[\protect\citeauthoryear{Sguera, Galeano, and Lillo}{Sguera
  et~al.}{2015}]{Sguera2015}
Sguera, C., P.~Galeano, and R.~E. Lillo (2015, 7).
\newblock Functional outlier detection by a local depth with application to nox
  levels.
\newblock {\em Stochastic Environmental Research and Risk Assessment\/}~{\em
  28\/}(462), 1835--1851.

\bibitem[\protect\citeauthoryear{Soueidatt}{Soueidatt}{2014}]{funclustR}
Soueidatt, M. (2014).
\newblock {\em Funclustering: A package for functional data clustering.}
\newblock R package version 1.0.1.

\end{thebibliography}
\end{document}